\documentclass[useAMS,usenatbib]{mn2e}

\usepackage{amssymb}
\usepackage{amsmath}
\usepackage{times}

\newcommand{\HI}{H\,{\sc i}}
\newcommand{\nhi}{\mbox{$N_{\rm HI}$}}
\newcommand{\msun}{\mbox{$M_\odot$}}
\newcommand{\mhi}{\mbox{$M_{\rm HI}$}}
\newcommand{\kms}{\mbox{km s$^{-1}$}}
\newcommand{\cm}{cm$^{-2}$}

\title[HI in Leo~T]
{The Local Group dwarf Leo~T: HI on the brink of star formation}

\author[Ryan-Weber et al.]{Emma V. Ryan-Weber,$^{1}$\thanks{email: eryan@ast.cam.ac.uk} 
Ayesha Begum$^1$, Tom Oosterloo$^{2,3}$, Sabyasachi Pal$^4$,\newauthor
Michael J. Irwin$^1$, Vasily Belokurov$^1$, N. Wyn Evans$^1$, and Daniel B. Zucker$^1$ \\ 
$^1$Institute of Astronomy, Madingley Rd, Cambridge, CB3 0HA, UK\\ 
$^2$Netherlands Foundation for Research in Astronomy, Postbus 2, 7990 AA Dwingeloo, 
The Netherlands\\
$^3$Kapteyn Astronomical Institute, University of Groningen, Postbus 800, 
9700 AV Groningen, The Netherlands\\
$^4$National Centre for Radio Astrophysics, Tata Institute of Fundamental Research, 
Pune 411-007, India}

\begin{document}

\date{Accepted 2007 November 15. Received 2007 November 12; in original form.2007 August 3}

\pagerange{\pageref{firstpage}--\pageref{lastpage}} \pubyear{2007}

\maketitle

\label{firstpage}

\begin{abstract}
  We present Giant Meterwave Radio Telescope (GMRT) and Westerbork
  Synthesis Radio Telescope (WSRT) observations of the recently
  discovered Local Group dwarf galaxy, Leo~T. The peak \HI\ column
  density is measured to be $7\times10^{20}$ \cm, and the total \HI\
  mass is $2.8\times10^5$ \msun, based on a distance of 420 kpc. Leo~T
  has both cold ($\sim500$ K) and warm ($\sim 6000$ K) \HI\ at its
  core, with a global velocity dispersion of 6.9 \kms, from which we
  derive a dynamical mass within the \HI\ radius of $3.3\times10^6$
  \msun, and a mass-to-light ratio of greater than 50. We calculate
  the Jeans mass from the radial profiles of the \HI\ column density
  and velocity dispersion, and predict that the gas should be globally
  stable against star formation. This finding is inconsistent with the
  half light radius of Leo~T, which extends to 170 pc, and indicates
  that local conditions must determine where star formation takes
  place. Leo~T is not only the lowest luminosity galaxy with on-going
  star formation discovered to date, it is also the most dark matter
  dominated, gas-rich dwarf in the Local Group.

\end{abstract}


\section{Introduction}
\label{sec:intro}

Leo~T is an impressively small, yet complex dwarf galaxy. A member of
the Local Group, Leo~T is the lowest luminosity galaxy discovered
to date with on-going star formation (Irwin et
al. \citeyear{Irwin07}). Its colour-magnitude diagram reveals both a
red giant branch and young blue stars, $\sim6-8$ Gyrs and $\sim200$
Myr in age, respectively. Although its stellar morphology and
intermediate-aged red stars are similar to the dwarf spheroidal (dSph)
galaxies, many of which have been discovered recently (Belokurov et
al. \citeyear{Belokurov07}, and references within), the presence of a
younger, blue stellar population is more typical of a dwarf irregular
(dIrr) galaxy. This duality has lead to the 'transition' label, hence
the name Leo~T. In addition, dSph galaxies are usually found within
250 kpc of Milky Way, whereas Leo~T is located at a distance of 420
kpc, similar to that of other transitional dwarfs, such as Phoenix,
and dIrr galaxies \citep{Grebel00}. The presence of cool gas is
another trademark of transitional dwarfs and dIrrs. As reported in
Irwin et al. (\citeyear{Irwin07}), Leo~T has a spatially coincident
detection of \HI\ in the Northern \HI\ Parkes All Sky Survey (HIPASS,
Wong et al. \citeyear{Wong06}). Recent optical spectroscopy has
confirmed that the stellar recessional velocity matches the
\HI\ radial velocity measurement \citep{Simon07}.

The study of the smallest dwarf galaxies provides insight into how the
least massive dark matter haloes retain cool gas and form stars. The
number of dark haloes in the Local Group predicted by cosmological
simulations is typically of the order of hundreds \citep{Klypin99,
  Moore99}. Only those dark haloes that can maintain sufficient cool
gas, allowing star formation to proceed, produce luminous dwarf
galaxies. The various processes that are thought to suppress star
formation in low mass haloes include a change in the Jeans mass due to
global reionization \citep[e.g.,][]{Efstathiou92, Benson02, Cooray05},
the heating and removal of gas via supernovae and stellar wind feedback
\citep[e.g.][]{Dekel03, Ricotti05}, ram pressure and tidal stripping
\citep[e.g.][]{Blitz00}, or simply a temperature floor in the
interstellar medium \citep{Kaufmann07}. Mass-to-light measurements of
Local Group dwarfs suggest that each galaxy is embedded in a dark
matter halo with a mass of about $10^7$ \msun\ \citep{Mateo98}. The
idea that there is a minimum dark matter halo mass able to form stars
is supported by analytic calculations \citep{Taylor05}, observational
data \citep{Gilmore07} and numerical simulations \citep{Ricotti05,
  Read06}. Thus observations of the presence, morphology and
kinematics of cool gas in dwarf galaxies provide important constraints
on these predictions and hold ramifications for our understanding of
galaxy formation and evolution. Leo~T is a crucial piece of evidence
in this line of inquiry, as it is the faintest dwarf galaxy detected
to date with on-going star formation.

In this letter we present Giant Meterwave Radio Telescope (GMRT) and
Westerbork Synthesis Radio Telescope (WSRT) observations that confirm
the presence of \HI\ in Leo~T at higher spatial and velocity
resolution than the initial detection in HIPASS. In section~3 we give
the \HI\ parameters of Leo~T, and calculate the Jeans mass profile. A
discussion of how star formation has proceeded in Leo~T and a
comparison to other Local Group dwarfs is given in section~4.

\section{Observations and Data reduction}
\label{sec:data}

\subsection{GMRT data}

The GMRT \citep{Swarup91} observations of Leo~T (centred on RA (2000):
09$^h$34$^m$53.5$^s$, DEC(2000): $+{17}^{\circ} 02^\prime
52.0^{\prime\prime}$ ) were conducted on December 12 2006. An
observing bandwidth of 1 MHz centered at 1420.36 MHz (which
corresponds to a heliocentric velocity of 35 \kms) was used. The band
was divided into 128 spectral channels, giving a channel spacing of
1.65 \kms. Absolute flux and bandpass calibration was done using scans
on the standard calibrator 3C286, which were observed at the start and
end of the observing run. Phase calibration was done using the VLA
calibrator 0842+185, which was observed once every 40 minutes. The
total on-source time was $\sim$ 5 hours.

The data were reduced using standard tasks in classic
AIPS\footnote{Astronomical Image Processing System}. For each run, bad
visibility points were edited out, after which the data were
calibrated. A low resolution data cube ($39\arcsec\times47\arcsec$)
was made using the AIPS task IMAGR. The RMS noise per channel at this
resolution is 5.2~mJy beam$^{-1}$. The \HI\ emission from Leo~T
spanned 7 channels of the spectral cube. A continuum image was also
made using the average of remaining line free channels. No continuum
was detected at the location of Leo~T to a $3\sigma$ flux limit of
2.3~mJy beam$^{-1}$ (for a beam size of $39\arcsec\times47\arcsec$).

Moment maps were made from the data cube using the AIPS task MOMNT. To
obtain the moment maps, lines of sight with a low signal to noise
ratio were excluded by applying a cutoff at the $2\sigma$ level
($\sigma$ being the rms noise level in a line free channel), after
smoothing in velocity (using boxcar smoothing three channels wide) and
position (using a Gaussian with a FWHM approximately twice that of the
synthesized beam). A map of the velocity dispersion was also made in
GIPSY\footnote{Groningen Image Processing SYstem} using single
Gaussian fits to the individual profiles. From the Gaussian fits, we
find a median velocity dispersion $\sigma_v\sim$ 3 \kms.

%% Figure 1
\begin{figure}	
 \vspace{8cm} 
 \includegraphics{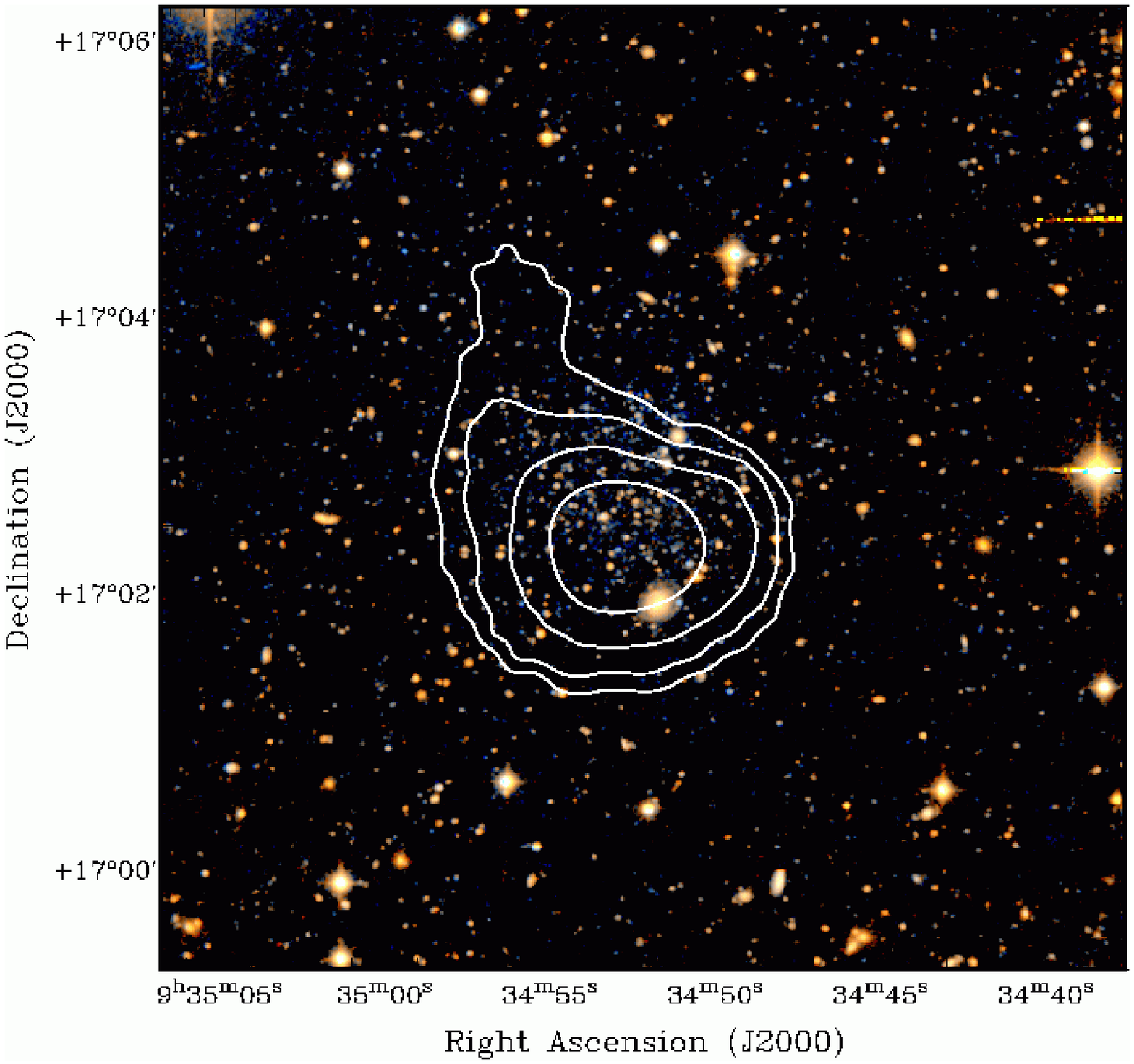} 
\caption{Colour image of Leo~T from INT WFC $g$ and $r$ band data with
  GMRT \HI\ contours overlaid.  The column density contours at
  2, 5, 10 and 20x10$^{19}$ \cm, and the beam size is 
$39\arcsec\times47\arcsec$.}
\label{fig:GMRTolay}
\end{figure}

%% Figure 2
\begin{figure}	
 \vspace{8cm} 
 \includegraphics{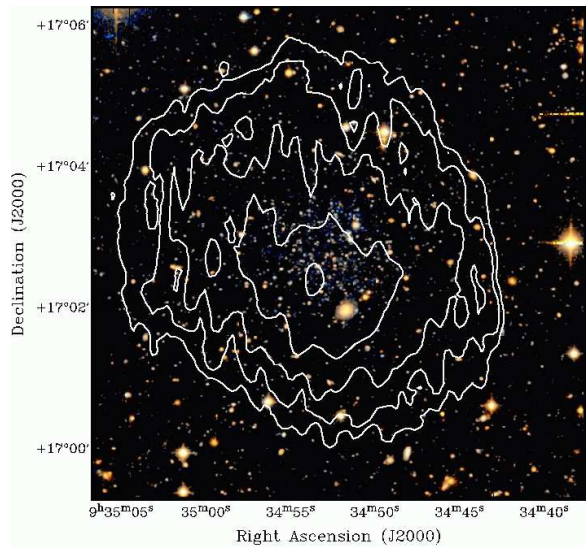} 
\caption{The same colour image from Figure~\ref{fig:GMRTolay}, however
  the \HI\ contours are from the WSRT data. In this case, the column
  density contours 2, 5, 10, 20 and 50x10$^{19}$ cm$^{-2}$, and the
  beam size is $12.9\arcsec\times50.4\arcsec$. An explanation for the
  difference between the GMRT and WSRT detections is given in
  Section~\ref{sec:results}.}
\label{fig:WSRTolay}
\end{figure}

\subsection{WSRT data}

Leo~T was observed with the WSRT on the night of February 16/17,
2007. The observations consisted of a full 12-hr synthesis, using the
so-called maxi-short configuration that gives good coverage of the
inner $uv$ plane, although, given the low declination of Leo~T, data
collected at extreme hour angles had to be flagged because of
shadowing on the shortest baselines. The observing bandwidth was 5 MHz
(corresponding to about 1000 \kms), using 1024 channels with 2
independent polarisations.  The data processing was done using the
MIRIAD package \citep{sault95}. Before and after the 12-hr observation,
a standard WSRT calibrator was observed (3C147 and CTD93), from which
the spectral response of the telescope was determined.  As is standard
practice with the WSRT, during the 12-hr track no additional (phase)
calibrators were observed. Instead, the large bandwidth allows us to
determine the gain variations by self-calibration of the continuum
image made from the line-free channels of the data.

Two datacubes were made, with spatial resolutions
$12.9^{\prime\prime}\times 50.4^{\prime\prime}$ and $
28.2^{\prime\prime} \times 45.2^{\prime\prime}$, respectively. Each
datacube was gridded into 900 channels 1.0 \kms\ wide to which
additional Hanning smoothing was applied, for a resulting velocity
resolution of 2.0 \kms. The datacubes were cleaned using the
\cite{Clark80} algorithm. In an iterative procedure, regions with line
emission were identified by smoothing the data to twice the spatial
resolution and selecting a clip level by eye to define the masked
region where the data were cleaned. This procedure was repeated until
convergence was achieved. The final noise in the Hanning-smoothed data
is 1.3 and 2.0 mJy/beam for the high and low resolution datacubes
respectively.

To construct the total \HI\ images and the velocity fields, the same
masks were used. The \HI\ flux integral is $6.7$ Jy \kms,
corresponding to a \HI\ mass of $2.8 \times 10^5$ \msun (using a
distance to Leo~T of 420 kpc). The mass derived here -- given the
uncertainties of the original detection -- is consistent with that
derived from the HIPASS data ($2 \times 10^5$ \msun).  No continuum
source associated with Leo T was detected to a $3\sigma$ flux limit of
0.5~mJy beam$^{-1}$.

\section{HI Properties}
\label{sec:results}

\begin{table}
\begin{minipage}{100mm}
\caption{Measured and derived properties of Leo~T from \HI\ data}
  \begin{tabular}{ll} \hline Parameter&\\ 
  \hline 
Optical coordinates (J2000)        & 09:34:53.4 +17:03:05\\
\HI\ centre                & 09:34:53.5 +17:02:22\\
\HI\ radius, $r_{\rm HI}$   & 2.5\arcmin\ (300 pc)\\ 
$S_{\rm int}$               & 6.7 Jy \kms\\ 
$\mhi$                     & $2.8\times10^5$ \msun \\
$\nhi(\rm peak)$           & $7\times10^{20}$ \cm \\
$v_\odot$                   & 38.6 \kms\\
$\sigma_v$(CNM $T\sim500$ K) & 2 \kms\\
$\sigma_v$(WNM $T\sim6000$ K)& 7 \kms\\
$\sigma_v$(global profile)   & 6.9 \kms\\
M$_{\rm dyn}$               & $>3.3\times10^6$ \msun \\
$r_{\rm HI}/r_{\rm Plummer}$  & 1.8\\
M$_{\rm dyn}$/L$_V$         & $>56$ \\ 
f$_{\rm gas}$ = M$_{\rm gas}$/(M$_{\rm gas}$ + M$_{\rm star}$)& 0.8 \\
f$_{\rm baryonic}$ = (M$_{\rm gas}$ + M$_{\rm star}$)/M$_{\rm dyn}$& 0.15\\
  \hline
\end{tabular}
\end{minipage}
\end{table}

The GMRT observations detect a central, cool core of \HI\ (see
Figure~\ref{fig:GMRTolay}), as evidenced from the low median velocity
dispersion of 3 \kms. The mass of \HI\ contained in this cool
component is $1.2\times10^5$ \msun, which is approximately 40 per cent
of the total \HI\ mass of Leo~T. At the resolution of the images presented
in Figures~\ref{fig:GMRTolay} and \ref{fig:WSRTolay} there is a factor
of four difference in sensitivity. Formally the GMRT data have a 1
channel 3$\sigma$ column density sensitivity of $2\times10^{19}$ \cm.
Comparing the contours in Figures~\ref{fig:GMRTolay} and
\ref{fig:WSRTolay} it is obvious that despite the formal sensitivity,
the GMRT observations miss a substantial fraction of the flux in the
outer parts of Leo~T. Due to GMRT's lack of short baselines, the GMRT
observations fail to detect the warmer component of the \HI, which is
spread over a larger spatial scale. The WSRT data detect \HI\ out to
a radius of 2.5\arcmin, or 300 pc (using a distance of 420 kpc).  We
are confident that this is the true extent of \HI\ in Leo~T, as the
entire HIPASS flux density is recovered by the WSRT observations. The
centre of the \HI\ detection is 40\arcsec\ to the south of the stellar
centre of Leo~T. This offset is less than the extent of the beam major
axis for both the GMRT and WSRT observations, although it is
interesting to note that the same offset is found in the two
independent data sets.

\begin{figure}	
 \vspace{6cm} 
 \includegraphics{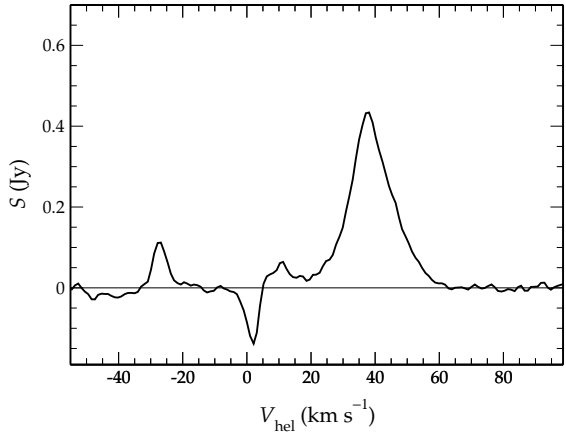} 
\caption{The global \HI\ spectrum of Leo~T from the WSRT data in the
  heliocentric frame. Emission at less than 20 \kms\ does not 
  appear to be physically connected to Leo~T. The negative feature is
  poorly imaged Galactic \HI\ due to the lack of short baselines.}
\label{fig:spectrum}
\end{figure}

\begin{figure}	
 \vspace{7cm} 
 \includegraphics{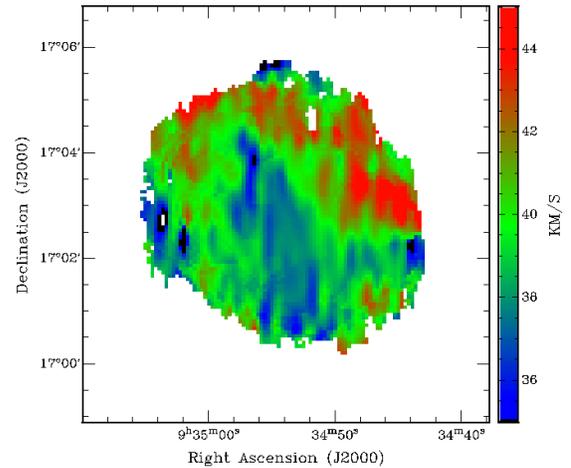} 
\caption{The velocity field of Leo~T derived from WSRT data; the field is
 chaotic, with little evidence for rotation.}
\label{fig:velfield}
\end{figure}

\begin{figure}	
 \vspace{6cm} 
 \includegraphics{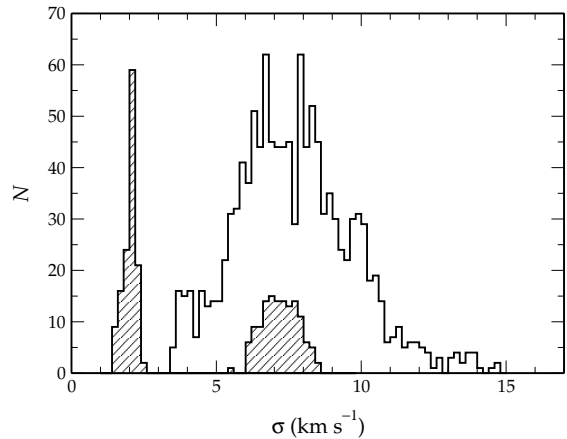} 
\caption{Histogram of \HI\ velocity dispersions from the WSRT data.
  The unshaded histogram gives the distribution of $\sigma_v$ values
  from fitting a single Gaussian to the velocity profile at each
  spatial pixel in the data cube. In the central bright region, these
  fits have significant systematic residuals, thus double Gaussian
  fits to pixels in this region are warranted. The shaded histogram
  gives the distribution of $\sigma_v$ values for a double Gaussian fit
  to velocity profiles in the bright central region.}
\label{fig:sigma}
\end{figure}

Figure~\ref{fig:spectrum} shows the global \HI\ profile from the WSRT
data. The global kinematic properties of the \HI\ are in good
agreement with the stars, which have a mean velocity of 38.1 \kms\ and
a velocity dispersion of 7.5 \kms\ \citep{Simon07}. Leo~T has a
chaotic velocity field, with some evidence of a gradient, but no
indication of systematic rotation (see
Figure~\ref{fig:velfield}). This type of velocity field is typical of
the \HI\ in low mass dIrr galaxies that have been observed with
sufficient velocity resolution, for example, LSG-3
\citep{Young97}. Interestingly, recent results from FIGGS (Faint
Irregular Galaxies GMRT Survey) have now shown that most dwarf
galaxies below a dynamical mass of $\sim10^{8}$ \msun, show velocity
fields that are either completely chaotic or show large scale patterns
that are not consistent with systematic rotation (Begum et al., in
preparation). We measure a global \HI\ velocity dispersion of 6.9
\kms, although the GMRT data indicate that there is a central
component of cold gas with a much lower
dispersion. Figure~\ref{fig:sigma} gives the velocity dispersion
histogram across the whole face of Leo~T using the WSRT data, where a
single Gaussian is fit to each pixel in the data cube, and the
$\sigma_v$ recorded. A second histogram shows the results of a double
Gaussian fit to each pixel in the bright, central region of Leo~T. The
double Gaussian describes the shape of the spectra substantially
better in the central region, as evidenced from the significant
residuals that result from fitting a single Gaussian. The double
Gaussian fits show that gas in the centre of Leo~T consists of cold (2
\kms, $T\sim500 K$) and warm (7 km/s, $T\sim6000 K$) components. This
two-phased medium is typical of faint dIrr galaxies; double-Gaussian
line profile fits of a sample of ten gas-rich dwarf galaxies
\citep{Begum06} show narrow components of the velocity dispersion in
the range $2-7$ \kms, while broad components range in value from 6 to
17 \kms. Some Local Group galaxies also exhibit a similar two-phased
interstellar medium, for example, Leo~A \citep{Young96} and Sag DIG
\citep{Young97}. By contrast the Local Group dwarf LGS~3 - which has
  ceased forming stars - lacks a cold \HI\ phase \citep{Young97}. In
    Leo~T, the single Gaussian fit pixels in all regions give velocity
    dispersions ranging from 3 to 15 \kms, with a median of 7.8
    \kms. The high dispersion tail ($11-15$ \kms) corresponds to a
    region 30\arcsec\ North and 70\arcsec\ East of the \HI\ centre.

Given the lack of any systematic rotation, it is difficult to
accurately determine the total dynamical mass for Leo~T. Assuming the
system is in equilibrium and that the \HI\ dynamics are a fair tracer
of the overall mass distribution the virial theorem can be used to
determine the mass. Applying the virial theorem under the assumption
of a spherical \HI\ distribution and an isotropic velocity dispersion
with negligible rotation, the dynamical mass is given by M$_{\rm
  dyn}=r_{\rm g}\sigma_v^{2}/G$, where $r_{\rm g}$ is the
gravitational radius, $\gtrsim 300 $ pc, and $\sigma_v=6.9$ \kms\ from
the global profile. Thus, the total dynamical mass is M$_{\rm dyn}
\gtrsim 3.3\times10^{6}$\msun. This is a lower limit to the total
mass, as the dark matter may be more extended than the
\HI. \cite{Simon07} estimate a total mass of $7.3\times10^6$ \msun,
using the stellar velocity dispersion and a method that accounts for a
more extended gravitational radius based on the Plummer profile. Leo~T
is quite gas rich, with more mass contained in \HI\ than in
stars. Specifically, given its luminosity L$_V=6\times10^4$\msun, then
assuming a conservative stellar mass-to-light ratio of 2, we obtain a
stellar mass of M$_{\rm star} \sim 1.2\times10^5$\msun. If we correct
for Helium and metals (but not molecular gas) the gas fraction of
Leo~T is 80 per cent. Given the extremely low stellar mass, the
observed mass-to-light of Leo~T is quite high, M$_{\rm dyn}$/L$_V
\gtrsim 56$.

\begin{figure}	
 \vspace{6cm} 
 \includegraphics{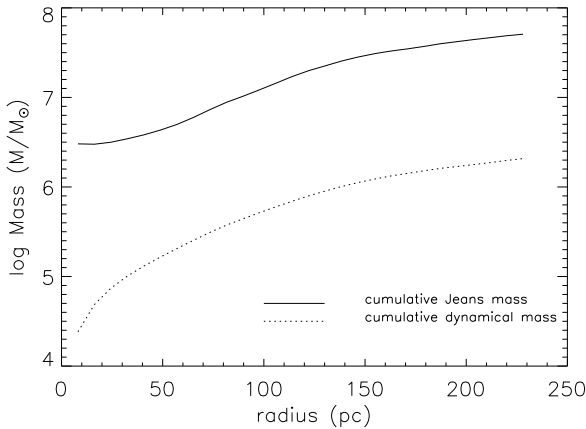} 
\caption{Cumulative Jeans mass (solid line) and dynamical mass profile
  (dotted line) of Leo~T. The Jeans mass profile is derived from the
  radially averaged column density and velocity dispersion (see
  Equation 1 and 2).}
\label{fig:jeans}
\end{figure}

To explore the relationship between the \HI\ properties and the location
of stars in Leo~T we calculate the differential Jeans mass profile.
For a spherically symmetric system, with uniform density
$\langle \rho \rangle$, the Jeans mass is given by
\begin{equation}
M_{\rm Jeans}=\frac{1}{6}\pi\langle\rho\rangle\left(\frac{\pi c_s^2}{G\langle\rho\rangle}\right)^{3/2}
\end{equation}
\citep[Eqn 5-24,][]{Binney87}. In a gaseous system, the sound speed
$c_s$ is related to the velocity dispersion, $\sigma_v$, by
$c_s=\sigma_v/\sqrt{5/3}$ \citep[e.g.][]{Schaye04}. The assumption of
homogeneity in Equation 1 only affects the timescale of the collapse,
not the stability condition itself \citep{Penston69}.  Thus, we can
use the measured density and velocity dispersion profiles to calculate
the cumulative Jean mass. The \HI\ column density is related to the
three dimensional density $\rho$ by
\begin{equation}
\rho(r) = -\frac{m_p}{\pi} \int_r^\infty \frac{d\nhi}{dR}\frac{dR}{\sqrt{R^2-r^2}}
\end{equation}
\citep[Eqn 4-58,][]{Binney87}. The radially averaged velocity
dispersion smooths over the features seen in Figure~\ref{fig:sigma},
although a radial trend still exists, with a mean $\sigma_v$ of 3.5
\kms\ in the centre, increasing to 7 \kms\ at 230 pc. The velocity
dispersion profile is only calculated to a radius of 230 pc, as noise
in the data leads to a number of unreliable $\sigma_v$ measurements in
individual pixels in the outskirts of Leo~T. Therefore we have only
calculated the Jeans mass to this radius. The falling \HI\ column
density and increasing mean velocity dispersion cause the differential
Jeans mass to increase with radius, as shown in
Figure~\ref{fig:jeans}. The dynamical mass enclosed at radius $r$,
M$_{\rm dyn}(r)$, does not exceed the Jeans mass at any radius.

%% Figure 3

%% Figure 4

%% Figure 5

%% Figure 6

%% Figure 7
\begin{figure}	
 \vspace{6cm} 
 \includegraphics{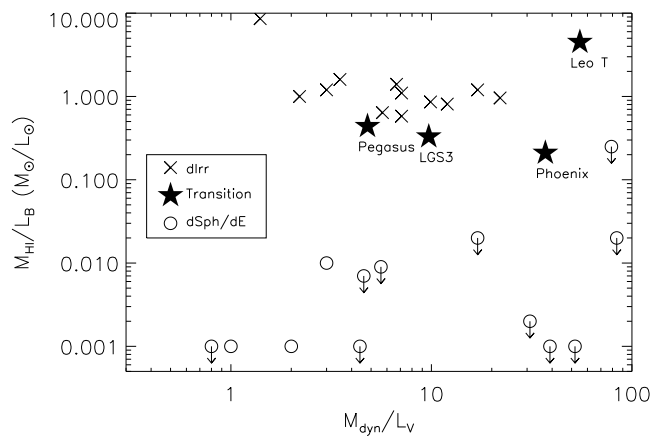} 
\caption{Comparison of mass-to-light properties of Leo~T with Local
  Group dwarf galaxies, where both the dynamical mass and at least a
  limit on the \HI\ is known. The \mhi/$L_B$ and total mass-to-light
  ratios are taken from Mateo (1998), Table~4. Arrows indicate upper
  limits to the \HI\ mass measurements.}
\label{fig:mass}
\end{figure}

\section{Star formation in a low mass halo}
\label{sec:discussion}

Given its stellar mass of $\sim1.2\times10^{5}$\msun\ and age of $\sim
6-8$ Gyrs, on average, Leo~T has been forming stars at the slow rate
of $1.5-2\times10^{-5}$\msun\ per year. Evidently this gentle star
formation rate has neither heated nor blown out all the gas in Leo~T,
allowing the stellar mass to build. A reservoir of approximately
$10^5$\msun\ of cool gas ($\sigma_v=2$ \kms, $T\sim 500 K$) remains,
which would take about $5\times10^{9}$ years to exhaust at the average
past star formation rate. Since Leo~T appears to be on such a gentle
trajectory with respect to the Galactic Standard of Rest (-58 \kms)
and Local Group (-97 \kms), it is possible that its gas has avoided
being stripped or heated by an interaction. Such a low central
velocity dispersion ($\sigma_v=2$ \kms) may be the reason why this
galaxy has been able to form stars despite its peak \HI\ column
density ($7\times10^{20}$\cm) being lower than those of typical low
surface brightness dwarf galaxies \citep{vanzee97}.

The Plummer law fit gives a stellar radius of 1.4\arcmin\ (Irwin et
al. \citeyear{Irwin07}), corresponding to a physical length of 170 pc.
An excess of stars above the background is detected out to a radius of
5\arcmin. Visual inspection of the image in Figures~\ref{fig:GMRTolay}
and \ref{fig:WSRTolay} shows that the over density of blue stars lies
within 1\arcmin, or 120 pc of the centre. The Jeans mass analysis
suggests that the gas is stable against collapse, contrary to what is
observed. This prediction is based the assumption of a spherical
distribution and radially averaged \HI\ column density and velocity
dispersion. Flattening the density distribution causes the Jeans mass
decrease. Since Leo~T appears to be stable against star formation
globally, it must be local processes that govern the star
formation. The tendency for locations with narrow velocity components
to coincide with \HI\ column densities above $3\times 10^{20}$ \cm\
supports this idea. Furthermore, a small level of star formation may
increase the turbulence, inhibiting the global collapse, but promoting
local instabilities via density fluctuations \citep{Schaye04}.

It would be interesting to repeat the radial Jeans mass analysis on
compact high velocity clouds (HVCs) to see whether their \HI\
properties suggest that star formation is possible. The analysis would
be straightforward since column density and velocity dispersion
measurement are independent of distance, an issue that has plagued the
interpretation of HVCs. Although Leo~T is more compact than any of the
compact HVCs for which high resolution imaging is available
\citep{Braun00}, their peak column densities ($\sim4\times10^{20}$\cm)
and minimum velocity dispersions ($<1$ \kms) are similar to those of
Leo~T. The initial detection of an \HI\ cloud at the location of Leo~T
in the HIPASS data was indistinguishable from an HVC. It is only the
high resolution \HI\ imaging presented here, together with a
recessional velocity measurement of the stars in Leo~T that has lead
to the confirmation that the \HI\ detected is of Leo~T itself. Given
the similarity in \HI\ characteristics, it is quite possible that
compact HVCs represent a population of 'failed' galaxies that contain
gas and no stars, with masses akin to that of Leo~T. Confirming this
suggestion, however, will be quite difficult, as distance measurements
rely on the chance alignment with both a foreground and background object, 
and the sky area covered by compact HVCs is small.

Compared with other dwarf galaxies in the Local Group, Leo~T contains a
significant component of dark matter given its large gas fraction. In
Figure~\ref{fig:mass}, dIrr galaxies tend to populate the top left of
the diagram, characterized by high gas fractions (\mhi/$L_B>0.1$) and
typical total mass-to-light ratios less than 10. On the other hand,
the lower half of the diagram is occupied by dSph galaxies, which have
low gas fractions (mostly upper limits as no \HI\ is detected in many
cases) and total mass-to-light ratios ranging from 1 to 100. The three
other galaxies classified as transitional dwarfs have properties that
lie in between the dIrr and dSph galaxies in
Figure~\ref{fig:mass}. Leo~T appears to stand alone in the top right
of diagram, it is significantly more gas rich and also contains more
dark matter than other transitional dwarfs.

The number and nature of Local Group dwarf galaxies provide important
constraints on models of galaxy formation and evolution. How many
other extremely low-luminosity, gas-rich galaxies are yet to be found
in the Local Group?  The SDSS (Sloan Digital Sky Survey) has surveyed
one fifth of the sky, and has uncovered just one such gas-rich dwarf,
namely Leo~T. The completeness for similar objects over the SDSS
survey area is close to one (Koposov et al. \citeyear{Koposov07}),
thus only a handful of other discoveries of comparable luminosity are
expected in other directions, assuming an isotropic distribution. The
other avenue to explore is detection via the \HI\ emission
line. Compared with HIPASS, current surveys, such as the Arecibo
Galactic \HI\ Survey (Stanimirovic et al. \citeyear{Stanimirovic06})
will provide four-times better spatial resolution with 0.2
\kms\ velocity resolution, and the Galactic All Sky Survey
\citep[GASS,][]{McClure-Griffiths06} will provide a 20-fold
improvement in velocity resolution. Disentangling extragalactic
\HI\ (with a chaotic velocity field) from emission associated with the
Galaxy at low Galactocentric velocities will always remain an issue;
the only definitive test is a matching velocity for the stellar
component of the object.

\section{Summary}
\label{sec:summary}
We have confirmed the presence of $2.8\times10^5$ \msun\ of \HI\ in
the Local Group dwarf galaxy, Leo~T. The gas is essentially centred on
the stellar emission with a radial extent of 300 pc. The \HI\ consists
of both a cold and warm neutral medium, revealed by a double Gaussian
fit with $\sigma_v=2,7$ \kms\ to pixels in the central region; the
velocity dispersion of the global profile is 6.9 \kms. The
comparatively low stellar mass gives Leo~T a high gas fraction of 80
per cent. The \HI\ extent and velocity dispersion of gas have been
used to estimate a total dynamical mass for Leo~T, M$_{\rm
  dynamical}\gtrsim 3.3\times10^6$ \msun. This value for the total
mass is just lower than the fiducial $10^7$\msun, below which galaxies
are expected to be dark \citep[e.g.,][]{Read06}, due to supernova
feedback and reionization.  Although the simulations predict that most
gas-rich dwarf galaxies form in haloes with M $>10^8$\msun, there are
some rare examples of galaxies in simulations that have a dark matter
halo mass of $\sim10^7$\msun and a small baryon fraction that do form
at least some stars, albeit inefficiently \citep{Ricotti05} -- a
description which fits Leo~T. It is interesting to note that Leo~T has
a significant fraction of dark matter; given that most gas-rich dwarfs
(i.e. dIrr) have M$_{\rm dyn}$/L$_V$ values between 1 and 10, the
total mass of Leo~T would be expected be about $10^5$\msun, yet a mass
in excess of $10^{6}$\msun\ has been measured, thus upholding the idea
of a minimum dark matter halo mass for dwarf galaxies \citep{Mateo98,
  Gilmore07}. We have compared the cumulative dynamical and Jeans
masses to determine that the \HI\ in Leo~T is {\em{globally}} stable
against star formation, this is inconsistent with the observed
presence of young stars. From this we conclude that local rather than
global processes must be responsible for Leo~T's stellar
component. The very low past average star formation rate may be the
reason why such cold gas is able to reside in Leo~T.

\section*{Acknowledgments}
The observations at the GMRT were conducted as a part of Director's
Discretionary time. The GMRT is operated by the National Centerer for
Radio Astrophysics of the Tata Institute of Fundamental Research. The
WSRT is operated by the Netherlands Foundation for Research in
Astronomy (ASTRON) with the support from the Netherlands Foundation
for Scientific Research (NWO).

\bibliographystyle{mn2e}
\bibliography{mn-jour,leot}

\bsp

\label{lastpage}

\end{document}